\documentclass[conference]{IEEEtran}
\usepackage{times}

\usepackage[numbers]{natbib}
\usepackage{multicol}
\usepackage[bookmarks=true]{hyperref}
\usepackage{todonotes}
\usepackage{amsmath}
\usepackage{amsfonts}
\usepackage{amssymb}
\usepackage{soul}
\usepackage{algorithm}
\usepackage{algpseudocode}
\usepackage{booktabs}
\usepackage{tabularx}
\usepackage{multirow}
\usepackage{subcaption}

\usepackage{amsthm}

\newtheorem{theorem}{Theorem}

\newtheorem{definition}{Definition}
\newtheorem{assumption}{Assumption}
\newtheorem{remark}{Remark}

\begin{document}

\title{A Data-Driven Structural Decomposition of Dynamic Games via Best Response Maps}
% \title{A Structural Reduction of Dynamic Games for Interaction-Aware Planning}
% You will get a Paper-ID when submitting a pdf file to the conference system
\author{Mahdis Rabbani, Navid Mojahed, Shima Nazari}

%\author{\authorblockN{Michael Shell}
%\authorblockA{School of Electrical and\\Computer Engineering\\
%Georgia Institute of Technology\\
%Atlanta, Georgia 30332--0250\\
%Email: mshell@ece.gatech.edu}
%\and
%\authorblockA{Twentieth Century Fox\\
%\and
%\authorblockN{James Kirk\\ and Montgomery Scott}
%\authorblockA{Starfleet Academy\\
%San Francisco, California 96678-2391\\

% avoiding spaces at the end of the author lines is not a problem with
% conference papers because we don't use \thanks or \IEEEmembership

%Homer Simpson\authorrefmark{2},
%Montgomery Scott\authorrefmark{3} and
%Eldon Tyrell\authorrefmark{4}}
%Telephone: (800) 555--1212, Fax: (888) 555--1212}
%\authorblockA{\authorrefmark{4}Tyrell Inc., 123 Replicant Street, Los Angeles, California 90210--4321}}

\maketitle

\begin{abstract}
Dynamic games are powerful tools to model multi-agent decision making, yet computing Nash (generalized Nash) equilibria remains a central challenge in such settings. Complexity arises from tightly coupled optimality conditions, nested optimization structures, and poor numerical conditioning. Existing game-theoretic solvers address these challenges by directly solving the joint game, typically requiring explicit modeling of all agents’ objective functions and constraints, while learning-based approaches often decouple interaction through prediction or policy approximation, sacrificing equilibrium consistency.
This paper introduces a conceptually novel formulation for dynamic games by restructuring the equilibrium computation. Rather than solving a fully coupled game or decoupling agents through prediction or policy approximation, a data-driven structural reduction of the game is proposed that removes nested optimization layers and derivative coupling by embedding an offline-compiled best-response map as a feasibility constraint. Under standard regularity conditions, when the best-response operator is exact, any converged solution of the reduced problem corresponds to a local open-loop Nash (GNE) equilibrium of the original game; with a learned surrogate, the solution is approximately equilibrium-consistent up to the best-response approximation error.
The proposed formulation is supported by mathematical proofs, accompanying a large-scale Monte Carlo study in a two-player open-loop dynamic game motivated by the autonomous racing problem. Comparisons are made against state-of-the-art joint game solvers, and results are reported on solution quality, computational cost, constraint satisfaction, and best-response residual.
\end{abstract}

\IEEEpeerreviewmaketitle

\section{Introduction}
% \subsection{Motivation and Challenges}
Decision-making and planning for multi-agent robotic systems, such as autonomous vehicles and mobile robots, require reasoning over strategic interactions among multiple agents. Dynamic games provide a principled framework for modeling such interactions, with generalized Nash equilibria capturing mutually optimal behavior under coupled dynamics and constraints \citep{Noncooperative1999Tamer,standardGNEPSolution, pustilnik2025generalized}.

In practice, however, computing generalized Nash equilibria in dynamic games remains challenging \citep{zhu2024sequentialdgsqp,pustilnik2025generalized}. Equilibrium computation typically entails tightly coupled optimality conditions and constraints, either solved simultaneously via a joint Karush-Kuhn-Tucker (KKT) or Mixed Complementarity Problem (MCP) formulation or implicitly through nested optimizations\footnote{It is noteworthy that by nested optimization here we mean the simultaneous interconnected optimization style which is traditionally used for solving Nash equilibrium. It should not be confused with the leader-follower style problem.} \citep{le2022algames,zhu2024sequentialdgsqp,fridovich2020efficient}. These structural couplings lead to poor numerical conditioning and fragile convergence behavior, particularly in the nonlinear and constrained settings that commonly arise in robotic systems.

Two complementary viewpoints dominate practical solution methods. The first is joint-equilibrium solving, in which one directly solves the fully coupled game using game-theoretic Nonlinear Programming (NLP) or MCP machinery \cite{pustilnik2025generalized,zhu2024sequentialdgsqp,le2022algames}. The joint coupled problem exhibits increased dimensionality, leading to high computational complexity, as well as potential inconsistencies in the solution due to infeasibility and convergence issues \citep{zhu2024sequentialdgsqp,le2022algames}.

The second viewpoint is the \emph{fixed-point} characterization of Nash equilibria through best responses, which motivates iterative best response (IBR) and Gauss--Seidel-style schemes \citep{dockner2001coordinate,SchwagerGameTheroreticPlanning}: alternate between solving one agent's optimal control problem while holding the other fixed until a fixed point is reached. Conceptually, IBR avoids forming a joint coupled system, but it introduces a different computational bottleneck: multiple evaluations of the non-ego agents' best response, which typically requires solving a full constrained optimal control problem \citep{sin2020iterativeBRDrawback,le2022algames}. Furthermore, the convergence of the IBR method is not guaranteed in general nonlinear, constrained, nonconvex games \citep{le2022algames}. In addition, many modern
learning--in--the--loop approaches that embed best-response operators require differentiating through the response, introducing additional chain-rule coupling that can return Stackelberg equilibrium, rather than Nash \citep{Li2023SolvingSC}.
% \todo[inline]{Be careful with the phrasing around IBR convergence. As written, "IBR convergence is not guaranteed..." can read like a drawback that our method resolves, which we do not claim. Consider reframing this as a general limitation of nonlinear/nonconvex constrained games (applies broadly), and then emphasize our actual contribution: avoiding repeated online BR OCP solves and avoiding differentiation-through-BR/chain-rule coupling, rather than guaranteeing convergence. Optionally add a sentence clarifying that the proposed reduction improves conditioning/structure but does not provide global convergence guarantees in general.}

While principled, all the above approaches generally assume explicit access to all agents' objective functions and constraints. Such assumptions are reasonable in centrally designed systems, but often unrealistic in settings such as human--robot interaction \citep{Soltanian20251419} and non-cooperative multi-robot systems \citep{chandra2025multi}, where agents act according to private objectives and constraints \citep{liu2023learning,smith2025mutual}. Representative examples include autonomous driving in mixed traffic, mobile robots operating in crowded environments such as warehouses, and robotic manipulators interacting in shared workspaces.

To mitigate these challenges, a large body of work has explored learning- and prediction-based approaches that decouple interaction by estimating future agent behavior \citep{Wei2014CConstantSpeed} or policies from data \citep{lopez2026data,lidard2024KLGameBlending}. While these methods can improve computational tractability, they typically significantly relax or completely abandon explicit equilibrium reasoning, and thus do not guarantee that the resulting solutions satisfy Nash equilibrium conditions. This lack of equilibrium consistency can be problematic in safety-critical applications, where it is important to reason explicitly about strategic reactions of other agents.

\subsection{Technical Contribution}
This paper proposes a conceptually novel formulation that restructures equilibrium computation itself. We adopt the fixed-point best-response perspective as the motivating lens, but rather than evaluating all best responses online (as in IBR) or solving a fully coupled joint system, we compile the non-ego agent's best-response dependencies from historical interactions and enforce them in the online optimal planning problem as a feasibility constraint. Concretely, we represent an agent's best response by a map, learned/fit from offline interactions, and then impose best-response feasibility as an explicit equality constraint within the ego agent's decision problem. This transforms the equilibrium computation from a nested or alternating sequence of coupled Optimal Control Problem (OCP) solutions into a reduced problem that can be solved in a single shot using standard NLP or MCP solvers. Importantly, the resulting optimality conditions are written in terms of the original decision variables and do not require differentiating through the best-response operator, thereby avoiding derivative coupling through the response map.

The primary contributions of this paper are:
\begin{enumerate}
    \item A structural reduction of finite-horizon dynamic games that enforces best-response consistency via feasibility constraints, enabling equilibrium computation without online nested best-response solves.
    \item A theoretical result establishing that solutions of the reduced formulation correspond to Nash equilibria of the original game, up to approximation error induced by the data-driven best-response representation.
    \item An empirical evaluation in an autonomous racing benchmark, including large-scale Monte Carlo comparisons against state-of-the-art joint game solvers.
\end{enumerate}

To the best of our knowledge, existing learning-based interaction methods either plan against predicted behaviors without enforcing equilibrium conditions, or learn components around a coupled equilibrium solver. In contrast, our approach uses data-driven modeling specifically to restructure equilibrium computation by replacing an online best-response optimality block with an offline-compiled best-response feasibility constraint, for the first time.

\subsection{Related Work}
\label{sec:related_work}

We review prior work on (i) joint generalized Nash equilibrium (GNE) solvers for dynamic games in robotics, (ii) learning-based interaction modeling for motion planning, and (iii) best-response and fixed-point perspectives on game solution methods.

A large body of robotics literature models interaction via dynamic games and computes equilibria by directly solving the coupled problem. Iterative linear-quadratic approximations (e.g., iLQ-game style methods) and associated software frameworks are widely used for continuous-state differential games in robotics \citep{fridovich2020efficient}. While these methods provide principled game-theoretic planning, they rely on local approximations and iterative refinement, and can be sensitive to initialization and numerical conditioning in nonlinear settings \citep{fridovich2020efficient}. For constrained general-sum games, several solvers target open-loop GNE by solving coupled optimality conditions. \citet{le2022algames} proposed ALGAMES, an augmented-Lagrangian method for constrained dynamic games with receding-horizon multi-vehicle examples, and sequential quadratic programming approaches have been developed for autonomous driving and racing, including DGSQP \citep{zhu2024sequentialdgsqp}. These methods directly address constrained equilibria but typically remain high-dimensional and tightly coupled across agents, and assume explicit access to all agents' objectives and constraints.

Complementary work exploits additional structure to reduce computational burden. Dynamic potential game formulations identify conditions under which multi-agent trajectory optimization admits a potential function, enabling equilibrium computation via a single optimal control problem \citep{RAPID2023Mehr, Williams2023Distributed}. While powerful, such structure is restrictive and does not generally apply to competitive general-sum interactions.

To avoid solving coupled games online, many planning pipelines decouple interaction by predicting other agents' trajectories or policies from data and planning against these predictions \citep{pmlrv168espinoza22a}. Other works integrate differentiable game layers into learning objectives, e.g., by learning costs or game parameters from demonstrations while solving a coupled game in an outer loop \citep{liu2023learning}. Recent approaches also blend model-based game reasoning with data-driven priors, such as KL-regularized dynamic games that bias solutions toward reference policies \citep{lidard2024KLGameBlending}, or amortize computation by training components from offline dynamic-game solutions \citep{pmlrv283kim25a}. These methods improve tractability or adaptivity, but typically do not remove the fundamental coupling of joint equilibrium computation and may sacrifice equilibrium consistency due to approximation error and additional nonconvexity.

Incomplete-information formulations motivate learning or estimation when agents' objectives are not fully known, e.g., in linear incomplete-information differential games \citep{Soltanian20251419}. More broadly, Nash equilibria can be characterized as fixed points of best-response operators \citep{Noncooperative1999Tamer}, motivating iterative best-response and local-update algorithms \citep{sin2020iterativeBRDrawback}. In robotics, best-response-like iterations also appear in iterative LQ approximations and potential-game variants via repeated local optimal control subproblems \citep{Williams2023Distributed}. In constrained settings, differentiable trajectory-game layers can be embedded in learning loops, but doing so typically requires differentiating through a coupled equilibrium solver and inherits its conditioning and convergence sensitivities \citep{liu2023learning}.

These perspectives motivate the central distinction of this paper: rather than iterating best responses online or differentiating through a coupled equilibrium solver, we use a data-driven best-response surrogate to \emph{restructure} equilibrium computation and eliminate nested optimization and derivative coupling at runtime, while retaining Nash-equilibrium consistency up to approximation error.

\section{Problem Statement}
\label{sec:problem_statement}

Consider a two-player finite-horizon dynamic game in discrete time. For each player $i\in\mathcal{I}=\{1,2\}$, let $x_{i,k}\in\mathbb{R}^{n_i}$ and $u_{i,k}\in\mathbb{R}^{m_i}$ denote state and control at time $k\in\{0,\dots,N-1\}$. The dynamics of player $i$ are
\begin{equation}
x_{i,k+1} = f_i(x_{i,k},u_{i,k}), \qquad k=0,\dots,N-1,
\label{eq:dyn_problem}
\end{equation}
with given initial condition $x_{i,0}$. Each player minimizes
\begin{equation}
\begin{aligned}
J_i(Z_i,Z_{-i}) = \sum_{k=0}^{N-1}\ell_i(x_{1,k},&u_{1,k},x_{2,k},u_{2,k})\\
&+\ell_{i,N}(x_{1,N},x_{2,N}),
\end{aligned}
\label{eq:cost_problem}
\end{equation}
% \todo[inline]{Notation consistency: earlier $J_i$ is written as $J_i(U_i,U_{-i})$ but later the equilibrium is defined over $Z_i=(X_i,U_i)$. Consider rewriting the cost as $J_i(Z_i,Z_{-i})$ throughout (or explicitly state that $X_i$ is determined by $U_i$ and $x_{i,0}$ via \eqref{eq:dyn_problem}).}

subject to coupled constraints
\begin{equation}
g_i(X_1,U_1,X_2,U_2)\le 0,\qquad h_i(X_1,U_1,X_2,U_2)=0,
\label{eq:constraints_problem}
\end{equation}
where $X_i:=\{x_{i,0},\dots,x_{i,N}\}$ and $U_i:=\{u_{i,0},\dots,u_{i,N-1}\}$. The subscript $_{-i}$ indexes player $\mathcal{I} \setminus i$, and $\ell_i(\cdot)$ and $\ell_{i,N}(\cdot)$ denote the stage and terminal costs, respectively. Let $Z_i:=(X_i,U_i)$ denote Player $i$'s stacked trajectory decision variables.

% \begin{definition}[Open-loop Nash equilibrium]
\begin{definition}
\label{def:definition_nash_loc}
A pair $(Z_1^\star,Z_2^\star)$ is a local open-loop generalized Nash equilibrium if there exist
neighborhoods $\mathcal{N}_i$ of $Z_{i}^\star$ , $\forall i\in\{1,2\}$, such that
\begin{align}
J_i(Z_i^\star,Z_{-i}^\star) &\le J_i(Z_{i},Z_{-i}^\star),
&&\forall\, Z_i\in \mathcal{N}_i \cap \Omega_i(Z_{-i}^\star), \label{eq:local_ne_p1}
\end{align}
where $\Omega_i(Z_{-i})=\big\{Z_i\,|\,g_i(Z_i,Z_{-i})=0,\, g_i(Z_i,Z_{-i})\leq 0\big\}$ is the feasible set for $Z_i$ from \eqref{eq:dyn_problem}, \eqref{eq:constraints_problem}.
\end{definition}
% \todo[inline]{Definition~\ref{def:definition_nash_loc} has notation/feasibility issues. (i) Replace $\mathcal{N}_1$ with $\mathcal{N}_i$. (ii) In $\Omega_i(Z_{-i})$, use $h_i(Z_i,Z_{-i})=0$ and $g_i(Z_i,Z_{-i})\le 0$ (currently it incorrectly has $g_i(\cdot)=0$). (iii) Decide whether $\Omega_i$ should also include the dynamics constraints $x_{i,k+1}=f_i(x_{i,k},u_{i,k})$ and the fixed initial condition $x_{i,0}$ (and any box bounds). If yes, define $\Omega_i$ accordingly; if no, rephrase the last sentence so it does not claim $\Omega_i$ comes from \eqref{eq:dyn_problem}. (iv) Either write the local GNE inequalities explicitly for $i=1,2$ or state "for each $i\in\{1,2\}$" clearly.}

% An open-loop generalized Nash equilibrium is a pair $(U_1^\star,U_2^\star)$ such that neither player can reduce its objective by unilateral deviation:
% \begin{equation}
% \begin{aligned}
% J_i(Z_i^\star,Z_{-i}^\star)\le J_i(Z_i,Z_{-i}^\star),\quad \forall Z_i \in \Omega_i(Z_{-i}). \end{aligned}
% \label{eq:nash_def}
% \end{equation}

In general nonlinear and constrained settings, GNE computation induces tightly coupled optimality conditions. A standard approach is to solve both players' Karush-Kuhn-Tucker (KKT) conditions simultaneously \citep{standardGNEPSolution}, often written compactly as a coupled residual system (or generalized equation)
\begin{equation}
\mathcal{F}_{\mathrm{KKT}}(Z_1,Z_2,\Lambda_1,\Lambda_2)=0,
\label{eq:joint_kkt_compact}
\end{equation}
where $\Lambda_i$ stacks multipliers (and complementarity variables, if present). Solving \eqref{eq:joint_kkt_compact} requires explicit knowledge of all agents' objectives and constraints and can be numerically fragile due to strong coupling and active-set switching.

An alternative viewpoint characterizes Nash equilibria as fixed points of best responses. Let
\begin{equation}
\mathrm{BR}_{i}(Z_{-i})\in\arg\min_{Z_{i}\in \Omega_{i}(Z_{-i})} J_{i}(Z_i,Z_{-i}).
\label{eq:br2_def}
\end{equation}
Iterative best response (IBR) performs Gauss--Seidel updates
\begin{equation}
% Z_1^\star \in \mathrm{BR}_1(Z_2^\star),\qquad Z_2^\star \in \mathrm{BR}_2(Z_1^\star).
Z_2^{t+1} \in \mathrm{BR}_2(Z_1^t),\qquad Z_1^{t+1} \in \mathrm{BR}_1(Z_2^{t+1}),
\label{eq:ibr}
\end{equation}
over $t$ iterations, and bypasses solving \eqref{eq:joint_kkt_compact} directly. by repeatedly solving system of coupled optimal control problems online.

% [the exact same explanation as: Under standard regularity conditions, an open-loop Nash equilibrium satisfies the coupled optimality system \eqref{eq:joint_kkt_compact}. Joint solvers attack this system directly, while IBR attempts to reach a fixed point by repeatedly re-solving one player's problem exactly. Both mechanisms are incompatible with Assumption~\ref{assump:asym_info}, since they require online access to Player~2's objective/constraints or an online best-response solve.]

\begin{assumption}[Asymmetric information]
\label{assump:asym_info}
Player~1 has full access to its own dynamics, objective, and constraints.
Player~1 does not have access to Player~2's objective and constraints, and cannot solve Player~2's best-response problem online.
Instead, Player~1 can observe (or estimate) Player~2's state trajectory over the horizon.
\end{assumption}

% \paragraph{Notation and the asymmetric viewpoint used later.}
% We use $Z_i := (X_i,U_i)$ to .
% In Sec.~\ref{sec:methodology}, we develop an asymmetric equilibrium computation in which Player~1 retains a full optimization model, while Player~2 is represented through an explicit best-response relation
% $Z_2 = \mathcal{B}_2(\cdot)$ compiled offline. This will allow us to compute Nash-consistent solutions without requiring online access to Player~2's objective and constraints.

\section{Decomposition via Best-Response Map}
\label{sec:methodology}

We propose an asymmetric structural reduction of open-loop Nash equilibrium computation.
The novelty is not merely replacing an agent with a predictor; rather, we restructure the equilibrium conditions themselves.
Specifically, we pull Player~2's optimality block out of the coupled equilibrium system and replace it with an explicit best-response feasibility block, thereby avoiding (i) joint coupled solves \eqref{eq:joint_kkt_compact} and (ii) repeated Gauss--Seidel alternations \eqref{eq:ibr}. This asymmetric design matches the common partial information structure in robotics.

% Our approach modifies \eqref{eq:joint_kkt_compact} at the system level. Instead of enforcing Player~2's KKT conditions online, we enforce that Player~2's decision variables are consistent with an offline-compiled response operator. Concretely, we keep Player~1's Nash stationarity structure intact (partial optimality against a fixed $U_2$), and replace the Player~2 optimality block by a feasibility equation.

Our key idea is to eliminate Player~2's optimality block from the online solve.
We introduce an explicit best-response selection of $\mathrm{BR}_2(\cdot)$, i.e.,
$\mathcal{B}_2(Z_1)\in \mathrm{BR}_2(Z_1)$ for all $Z_1$ in the region of interest and set
\begin{equation}
Z_2 = \mathcal{B}_2(Z_1),
\label{eq:true_br_relation}
\end{equation}

Crucially, instead of alternating between Player~1 and Player~2 solves as in IBR, we solve a single reduced equilibrium system consisting of:
(i) Player~1's optimality conditions with Player~2 treated as fixed, and
(ii) a best-response feasibility constraint enforcing \eqref{eq:true_br_relation}.
This yields an asymmetric system
\begin{equation}
\mathcal{F}_{\mathrm{red}}(Z_1,\Lambda_1,Z_2) =
\begin{bmatrix}
\mathcal{F}_{1,\mathrm{KKT}}(Z_1,\Lambda_1|Z_2)\\
Z_2 - \mathcal{B}_2(Z_1)
\end{bmatrix}
=0.
\label{eq:reduced_system}
\end{equation}

We keep $Z_2$ as an explicit variable and enforce the best-response relation as a separate feasibility block.
Therefore, Player~1's stationarity conditions remain \emph{partial} with respect to $Z_1$ holding $Z_2$ fixed, which is exactly the Nash structure.
This avoids the total-derivative (chain-rule) coupling that would arise from eliminating $Z_2$ and minimizing $J_1(Z_1,\mathcal{B}_2(Z_1))$, which would instead resemble an anticipatory/leader--follower effect.

\subsection{Illustrative Example}
\label{sec:analytic_example}

We illustrate the structural reduction on a minimal one-step, two-player unconstrained game.
Let $x_1,x_2\in\mathbb{R}$ be current positions and $v_1,v_2\in\mathbb{R}$ be one-step velocities, with
\begin{equation}
x_1^{+}=x_1+\Delta t\,v_1,\qquad x_2^{+}=x_2+\Delta t\,v_2 .
\label{eq:toy_dynamics}
\end{equation}
Each agent tracks a goal $g_i$ and penalizes effort and a soft separation error
$e_c:=(x_1^{+}-x_2^{+})-d$. The costs are
\begin{align}
J_1(v_1,v_2) &= \tfrac12 q_1(x_1^{+}-g_1)^2+\tfrac12 r_1 v_1^2+\tfrac12 w e_c^2, \\
J_2(v_2,v_1) &= \tfrac12 q_2(x_2^{+}-g_2)^2+\tfrac12 r_2 v_2^2+\tfrac12 w e_c^2 .
\end{align}

Set $\Delta t=1$, $q_1=q_2=r_1=r_2=1$, $w=2$, $d=0.5$, $x_1=x_2=0$, $g_1=1$, $g_2=-1$.
Let $y:=(x_1-x_2)-d=-0.5$, so $e_c=y+(v_1-v_2)$.
The Nash equilibrium satisfies the first-order conditions
$\frac{\partial J_1}{\partial v_1}=0$, $\frac{\partial J_2}{\partial v_2}=0$, which reduce to
\begin{equation}
4v_1-2v_2=2,\qquad -2v_1+4v_2=-2,
\label{eq:toy_linear_system}
\end{equation}
yielding
\begin{equation}
v_1^\star=\tfrac13,\qquad v_2^\star=-\tfrac13 .
\label{eq:toy_ne_solution}
\end{equation}

Instead of solving \eqref{eq:toy_linear_system} jointly, we extract Player~2's best-response map from
$\frac{\partial J_2}{\partial v_2}=0$:
\begin{equation}
v_2=\mathcal{B}_2(v_1)=\tfrac{v_1-1}{2}.
\label{eq:toy_br2}
\end{equation}
The reduced asymmetric system becomes
\begin{equation}
\begin{bmatrix}
\frac{\partial J_1}{\partial v_1}(v_1,v_2)\\
v_2-\mathcal{B}_2(v_1)
\end{bmatrix}
=
\begin{bmatrix}
4v_1-2v_2-2\\
v_2-\tfrac{v_1-1}{2}
\end{bmatrix}=0,
\label{eq:example_reduced_system}
\end{equation}
whose solution matches \eqref{eq:toy_ne_solution}.

\subsection{Implementation}
\label{sec:mcp_impl}

We now expand \eqref{eq:reduced_system} for the general constrained game in Sec.~\ref{sec:problem_statement}.
Assume $Z_1$ is the stacked decision variables of Player~1.
In the proposed asymmetric formulation, Player~2's variables $Z_2$ are treated as parameters in Player~1's optimization, and are constrained separately through a best-response feasibility relation.

Define Player~1's Lagrangian
\begin{equation}
\begin{aligned}
    \mathcal{L}_1(Z_1,\lambda_1,\mu_1\,|\,Z_2) & := J_1(Z_1|Z_2) + \lambda_{11}^\top h_1(Z_1|Z_2) \\
    & + \lambda_{12}^\top (Z_1^+-f_1(Z_1))+ \mu_1^\top g_1(Z_1|Z_2),
\end{aligned}
\end{equation}
with the Lagrange multipliers $\lambda_1=[\lambda_{11}, \lambda_{12}]$ and $\mu_1\ge 0$.

The KKT conditions for Player~1, holding $Z_2$ fixed, are
\begin{subequations}\label{eq:KKT_P1_main}
\begin{align}
\nabla_{Z_1} \mathcal{L}_1(Z_1,\lambda_1,\mu_1\,|\,Z_2) &= 0,\\
Z_1^+-f_1(Z_1) = 0,  \quad h_1(Z_1|Z_2) &= 0,\\
g_1(Z_1|Z_2)&\le 0,\\
\mu_1 &\ge 0,\\
\mu_1 \odot g_1(Z_1|Z_2)&=0.
\end{align}
\end{subequations}

We enforce Player~2 consistency through the best-response surrogate
\begin{equation}
r_{\mathrm{BR}}(Z_2,Z_1) := Z_2 - \mathcal{B}_2(Z_1)=0,
\label{eq:br_feasibility_general}
\end{equation}

Stacking \eqref{eq:KKT_P1_main} and \eqref{eq:br_feasibility_general} reduces the KKT system of equations \eqref{eq:reduced_system}, to
\begin{equation}
\mathcal{F}_{\mathrm{red}}(Z_1,\Lambda_1,Z_2) :=
\begin{bmatrix}
\nabla_{Z_1}\mathcal{L}_1(Z_1,\Lambda_1|Z_2)\\
Z_1^+-f_1(Z_1)\\
h_1(Z_1,Z_2)\\
g_1(Z_1,Z_2)\\
Z_2-\mathcal{B}_2(Z_1)
\end{bmatrix},
\label{eq:reduced_mcp_residual}
\end{equation}
together with bounds encoding $\mu_1\ge 0$ and any box constraints on $(Z_1,Z_2)$. Also, $\Lambda_1=\begin{bmatrix}\lambda_1,&\mu_1\end{bmatrix}$. 
This reduced KKT system can either be numerically solved by an NLP solver, like IPOPT, or can be reformulated as an MCP problem and solved by an appropriate solver, such as the PATH solver.

\subsection{Nash Interpretation}
\label{sec:nash_property}

% \begin{definition}
% \label{def:local_ol_ne}
% A pair $(Z_1^\star,Z_2^\star)$ is a \emph{local open-loop Nash equilibrium} if there exist
% neighborhoods $\mathcal{N}_1$ and $\mathcal{N}_2$ of $Z_1^\star$ and $Z_2^\star$, respectively, such that
% \begin{align}
% J_1(Z_1^\star,Z_2^\star) &\le J_1(Z_1,Z_2^\star),
% &&\forall\, Z_1\in \mathcal{N}_1 \cap \mathcal{F}_1(Z_2^\star), \label{eq:local_ne_p1}\\
% J_2(Z_2^\star,Z_1^\star) &\le J_2(Z_2,Z_1^\star),
% &&\forall\, Z_2\in \mathcal{N}_2 \cap \mathcal{F}_2(Z_1^\star), \label{eq:local_ne_p2}
% \end{align}
% where $\mathcal{F}_1(Z_2^\star)$ denotes Player~1's feasible set given $Z_2^\star$, and $\mathcal{F}_2(Z_1^\star)$ denotes Player~2's feasible set given $Z_1^\star$.
% \end{definition}

% \begin{assumption}
% \label{ass:p1_local_1}
% For a fixed $Z_2^\star$, Player~1's problem satisfies a constraint qualification at $Z_1^\star$
% (e.g., LICQ or MFCQ), so KKT conditions hold at $Z_1^\star$.
% Moreover, a second-order sufficient condition (SOSC) holds at $(Z_1^\star,\lambda_1^\star,\mu_1^\star)$,
% implying that $Z_1^\star$ is a (strict) local minimizer of Player~1's problem given $Z_2^\star$.
% \end{assumption}

% \begin{assumption}[Best-response map selects a local minimizer]
% \label{ass:p2_local_br}
% At $Z_1^\star$ (equivalently $Z_1^\star$), the embedded map satisfies
% $Z_2^\star=\mathcal{B}_2(Z_1^\star)$ and $Z_2^\star$ is a (strict) local minimizer of Player~2's problem
% given $Z_1^\star$.
% \end{assumption}

\begin{theorem}
\label{thm:reduced_is_local_nash}
% Suppose Assumptions~\ref{ass:p1_local_1} and \ref{ass:p2_local_br} hold.
Let $\Gamma^\star=(Z_1^\star,\Lambda_1^\star,Z_2^\star)$ be a solution of the reduced KKT
defined in~\eqref{eq:reduced_mcp_residual} together with the complementarity bounds.
Then $(Z_1^\star,Z_2^\star)$ is a local open-loop generalized Nash equilibrium.
\end{theorem}

\begin{proof}
Since $\Gamma^\star$ solves the reduced KKT, it holds the necessary condition of optimality at
$(Z_1^\star,\Lambda_1^\star)$ for the Player~1 problem with $Z_2$ treated as fixed at $Z_2^\star$. Thus,
\[J_1(Z_1^*,Z_2^*)\leq J_1(Z_1,Z_2^*), \qquad\forall Z_1\in\mathcal{N}_1\cap\Omega_1(Z_2^*).\]

 On the other hand, the best-response feasibility constraint holds:
\[
Z_2^\star=\mathcal{B}_2(Z_1^\star).
\]
Thus, $Z_2^*$ is a strict local minimizer of Player~2's optimization problem given a fixed $Z_1 = Z_1^*$. Hence,
\[
J_2(Z_2^*,Z_1^*) \leq J_2(Z_2,Z_1^*), \qquad \forall Z_2\in\mathcal{N}_2\cap \Omega_2(Z_1^*).
\]

The two inequalities above construct \eqref{eq:local_ne_p1} in the Definition~\ref{def:definition_nash_loc}. Thus, $(Z_1^\star,Z_2^\star)$ is a local open-loop generalized Nash equilibrium.
\end{proof}

%\label{sec:nash_property}

\section{Data-Driven Best-Response Operator}
\label{sec:br_operator}

The proposed reduction requires an explicit operator $\mathcal{B}_2$ that maps observable features of Player~1 to Player~2's best-response decision variables $Z_2$.
Under asymmetric information (Assumption~\ref{assump:asym_info}), we are able to construct a data-driven approximation $\widehat{\mathcal{B}}_2$ from interaction data to compensate for the partial information.
In many robotics settings, players lack access to others' private objective and constraint specifications. Instead, they may only observe surrounding players' state evolution (e.g., trajectories estimated from perception) and possibly some nominal motion model class. Thus, we make a realistic assumption of setting $Z_1$ to be Player~1's state trajectory $X_1$.
% \todo[inline]{Notation fix: do not write "set $Z_1$ to be $X_1$" since $Z_1=(X_1,U_1)$. Instead, introduce an information/feature variable, e.g., $\Xi_1 := X_1$ (or $\Xi_1=\Psi(X_1)$), and define the operator as $\mathcal{B}_2(\Xi_1)$. Update the section accordingly.}

\subsection{Data-Driven Approximation}
We fit $\widehat{\mathcal{B}}_2$ from samples $\{(X_1^{(j)},Z_2^{(j)})\}_{j=1}^M$, where $Z_2^{(j)}$ is obtained from demonstrations, logs, or offline best-response solves in simulation.
The reduced KKT then uses the feasibility constraint
\begin{equation}
Z_2 - \widehat{\mathcal{B}}_2(X_1)=0.
\label{eq:learned_br_constraint}
\end{equation}

\begin{remark}[Approximate equilibrium consistency]
When $\widehat{\mathcal{B}}_2$ is approximate, the reduced solve enforces a best-response residual
$\|r_{\mathrm{BR}}\| =\| Z_2-\widehat{\mathcal{B}}_2(X_1)\|\geq 0$.
The measure $\|r_{\mathrm{BR}}\|$ is an explicit measure of best-response inconsistency induced by approximation.
When $\widehat{\mathcal{B}}_2$ is exact, any solution of the reduced KKT satisfies the Nash equilibrium point in Definition~\eqref{def:definition_nash_loc}; when it is approximate, $\|r_{\mathrm{BR}}\|$ quantifies the violation of Player~2 best-response consistency.
\end{remark}

\begin{figure*}
    \centering
    \includegraphics[width=1\linewidth]{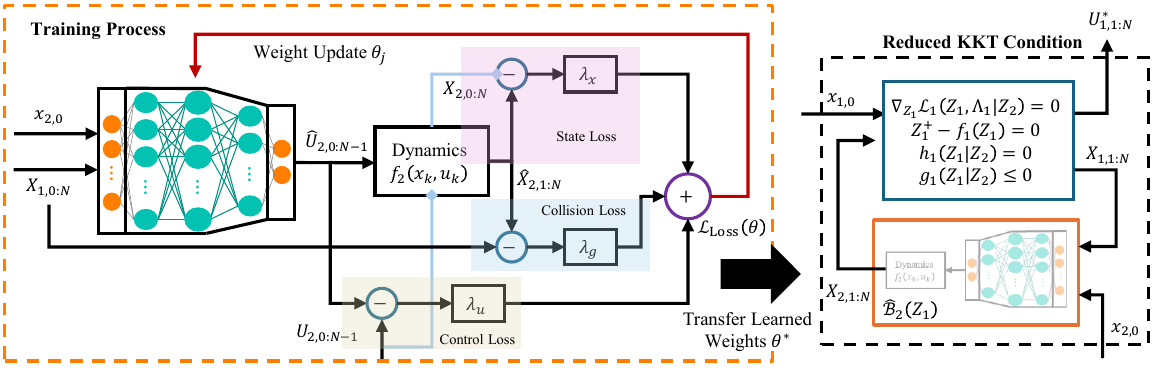}
    \caption{\textbf{Overall Pipeline}. During the training process on the left, the MLP learns the policy $\pi_\theta$ (see \eqref{sec:exp_br_mlp} through three different loss functions. Then, the best response approximation $\widehat{\mathcal{B}}_2$ is obtained by aggregating the learned policy and system dynamics of Player~2. Finally, we solve for the proposed reduced KKT Condition using the approximated best response surrogate. }
    \label{fig:pipeline}

\end{figure*}

Fig.~\ref{fig:pipeline} summarizes the overall pipeline of our proposed methodology, and our implementation for a benchmark problem of two-player racing scenario.

\section{Two--Player Racing Simulation}
\label{sec:experimental_eval}

This section evaluates the proposed methodology
Our goal is \emph{not} to introduce a new equilibrium solver that outperforms existing full-information solvers in speed or accuracy.
Rather, we demonstrate that:
\begin{enumerate}
    \item when Player~2's best-response operator is available (e.g., via offline computation), the reduced formulation recovers solutions consistent with the Nash definition in Sec.~\ref{sec:problem_statement}; and
    \item when Player~2's objective and constraints are not available online, a data-driven approximation $\widehat{\mathcal{B}}_2$ can be integrated through the feasibility constraint \eqref{eq:learned_br_constraint} to produce interaction plans that are safe, feasible, and approximately Nash-consistent.
\end{enumerate}
% \hl{Accordingly, beyond objective values and constraint satisfaction, we report the best-response residual
% $\|r_{\mathrm{BR}}\| := \|Z_2-\widehat{\mathcal{B}}_2(X_1)\|$
% as a direct measure of how closely the computed solution satisfies the embedded response relation.}

\subsection{Benchmark Problem: Two-Player Racing on a Constant--Curvature Track}
\label{sec:exp_benchmark}

We consider a two-player racing benchmark on a bounded constant--curvature (quarter--circle) track, which yields strong interaction due to limited lateral space and coupled collision-avoidance constraints.
Each agent is modeled by discrete-time kinematic bicycle dynamics in the track-aligned (Frenet) frame and solves a finite-horizon open-loop planning problem. 
Complete details of the Frenet dynamics, track parametrization, and bounds are provided in the Appendix.

For each player $i\in\{1,2\}$, the state and input are
\[
x_i = \begin{bmatrix} v_i & \psi_i & s_i & t_i \end{bmatrix}^\top, \qquad
u_i = \begin{bmatrix} a_i & \delta_i \end{bmatrix}^\top,
\]
where $s_i$ denotes progress along the centerline, $t_i$ denotes lateral deviation from the centerline, $v_i$ is speed, and $\psi_i$ is the heading (track-frame orientation) state.
The inputs are longitudinal acceleration $a_i$ and steering angle $\delta_i$.
Let $Z_i:=(X_i,U_i)$ denote the stacked trajectory variables over the horizon, consistent with Sec.~\ref{sec:methodology}.

% \paragraph{Ego planning problem (Player~1)}
% \label{sec:exp_p1}

Player~1 uses the full-information objective $J_1$ in \eqref{eq:cost_problem}.
We instantiate the stage and terminal terms as
\begin{equation}
\begin{aligned}
    &\ell_{1,k}
= R_u\|u_{1,k}\|_2^2
+ P_{\Delta u}\|u_{1,k}-u_{1,k-1}\|_2^2
+ q_v\,v_{1,k}^2,
\\
&\ell_{1,N}
= q_{12}s_{2,N}-q_{11}s_{1,N},
\end{aligned}
\label{eq:p1_stage_terminal}
\end{equation}
where $R_u$, $P_{\Delta u}$ are positive-definite matrices, $q_v$, $q_{11}$, and $q_{12}$ are positive weights, and $u_{1,-1} = 0$.
The terminal term promotes Player~1's progress while capturing competitive coupling through Player~2's terminal progress.

Given $x_{1,0}$ and a planning horizon $N=10$, Player~1 solves
\begin{equation}
\label{eq:Player1_optimization}
\begin{aligned}
\min_{\{u_{1,k}\}_{k=0}^{N-1}} \quad
& \ell_{1,N} + \sum_{k=0}^{N-1} \ell_{1,k}  \\
\text{s.t.}\quad
& x_{1,k+1}=f_1(x_{1,k},u_{1,k}), \qquad k=0,\dots,N-1,\\
& x_{\min}\le x_{1,k}\le x_{\max},\qquad u_{\min}\le u_{1,k}\le u_{\max},\\
& d_{\mathrm{safe}}^2-\|p_{1,k}-p_{2,k}\|_2^2 \le 0, \qquad k=0,\dots,N,
\end{aligned}
\end{equation}
where $p_{i,k}=[\mathbf{x}_{i,k},\,\mathbf{y}_{i,k}]^\top$ denotes the Cartesian position obtained from the Frenet state via the track mapping.
The optimization \eqref{eq:Player1_optimization} depends on Player~2 through $\{p_{2,k}\}$ and $s_{2,N}$, which in a standard dynamic game would be determined by Player~2's own objective and constraints.

% \subsection{Opponent model under asymmetric information (Player~2)}
% \label{sec:exp_p2}

Player~2 has its own objective and constraints (unknown to Player~1 online under Assumption~\ref{assump:asym_info}).
To close \eqref{eq:Player1_optimization} without querying Player~2's objective/constraints, we instantiate Player~2 through an offline-compiled best-response operator.
In particular, we approximate the mapping with a data-driven surrogate $\widehat{\mathcal{B}}_2$, and enforce it through the embedded feasibility constraint \eqref{eq:learned_br_constraint} as required by the reduced equilibrium system in Sec.~\ref{sec:mcp_impl}.
This preserves the Nash structure on the ego side (Player~1 remains optimal against a fixed $Z_2$) while removing any online dependence on Player~2's objective specification.

It is noteworthy that we assume Player~2's dynamics class is common knowledge and matches Player~1's kinematic model. As we want to have the dynamic feasibility for Player~2, we wrote the best response surrogate to predict the control input sequence, rather than the states, to make the dynamic feasibility explicit.
This is standard in autonomous driving and racing: vehicle motion is governed by universal physical constraints and can be identified from observed trajectories with high fidelity.
In contrast, objective functions and constraint preferences (e.g., aggressiveness, risk tolerance, racing style) are agent-specific and typically unobserved; this is precisely the asymmetry addressed by Assumption~\ref{assump:asym_info}.
% \todo[inline]{The reported best-response residual $\|r_{\mathrm{BR}}\|=\|Z_2-\widehat{\mathcal{B}}_2(X_1)\|$ is not very informative if the feasibility constraint \eqref{eq:learned_br_constraint} is enforced as a hard equality, since the solver will return $r_{\mathrm{BR}}\approx 0$ up to numerical tolerance. Consider instead reporting (i) a residual against a ground-truth/offline BR operator (e.g., $\|Z_2-\mathcal{B}_2(X_1)\|$ on held-out cases), or (ii) relaxing the constraint via a slack variable and penalizing $\|s\|$, then reporting $\|s\|$ as the BR inconsistency measure.}

% \todo[inline]{Clarify the output of the learned operator. The text says the surrogate predicts the control sequence for Player~2 to ensure dynamic feasibility, but \eqref{eq:learned_br_constraint} constrains $Z_2=(X_2,U_2)$. If $\widehat{\mathcal{B}}_2$ outputs only $U_2$, change the constraint to $U_2-\widehat{\mathcal{B}}_2(X_1)=0$ and explicitly include Player~2 dynamics as constraints to define $X_2$. Otherwise, state that $\widehat{\mathcal{B}}_2$ outputs the full $Z_2$.}
% \todo[inline]{The stage cost uses $\|u_{1,k}-u_{1,k-1}\|^2$, but $u_{1,-1}$ is undefined. Specify how $u_{1,-1}$ is handled (e.g., set to the previous applied control, or start this term from $k=1$, or define a fixed reference input for $k=0$).}

\subsection{Best-Response Surrogate: Architecture \& Training}
\label{sec:exp_br_mlp}

We implement $\widehat{\mathcal{B}}_2$ as a lightweight Multi-Layer Perceptron (MLP) that predicts Player~2's open-loop control sequence, and obtain Player~2's trajectory by rolling out the known dynamics.
As mentioned, this design guarantees dynamic feasibility of the predicted opponent motion by construction (with respect to $f_2$), and avoids treating the surrogate as a free-form state predictor.

% \paragraph{Parametrization.}
Let $\pi_\theta:\phi\to\hat{U}_{2,0:N-1}$ denote the trained network. For horizon $N$, we predict
\[
\widehat{U}_{2,0:N-1} = \pi_\theta(\phi),
\qquad
\widehat{U}_{2,0:N-1}\in\mathbb{R}^{N\times 2},
\]
where the feature vector $\phi$ summarizes the interaction context using:
(i) both players' initial states $(x_{1,0},x_{2,0})$, and
(ii) Player~1's planned state trajectory $X_{1,0:N}$ over the horizon.
Concretely,
\begin{equation}
\phi
=
\begin{bmatrix}
x_{2,0}\\
\mathrm{vec}(X_{1,0:N})
\end{bmatrix}.
\label{eq:phi_def}
\end{equation}

\begin{remark}
    This formulation does not violate causality. Online, $X_{1,0:N}$ is a decision variable in the reduced system, and the feasibility constraint \eqref{eq:learned_br_constraint} couples $Z_2$ to that decision variable through the fixed network parameters $\theta$.
\end{remark}

\paragraph{\textbf{Architecture \& Bounds.}}
We use a fully-connected MLP with $\tanh$ nonlinearities (hidden widths $128\!\rightarrow\!128\!\rightarrow\!64$) followed by a linear output layer.
To enforce actuator limits smoothly, we apply a differentiable squashing map per control dimension $d\in\{a,\delta\}$:
\begin{equation}
\hat{u}_{2,k}^{(d)} = m_d + h_d \tanh(\tilde{u}_{2,k}^{(d)}),
\label{eq:squashing}
\end{equation}
where $(m_d,h_d)$ are the midpoint and half-range determined by $(u_{\min},u_{\max})$.

Given $x_{2,0}$ and $\widehat{U}_{2,0:N-1}$, we compute Player~2's predicted trajectory via
\[
\hat{x}_{2,k+1}=f_2(\hat{x}_{2,k},\hat{u}_{2,k}),\qquad \hat{x}_{2,0}=x_{2,0}.
\]
We then construct $\widehat{Z}_2=(\widehat{X}_2,\widehat{U}_2)$ and use it in \eqref{eq:learned_br_constraint}.

\paragraph{\textbf{Training Loss.}}
We train $\pi_\theta$ using a rollout-aware objective that balances:
(i) control imitation, (ii) state consistency under rollout, and (iii) a soft collision penalty as follows,
\begin{equation}
\label{eq:br_loss}
\begin{aligned}
\mathcal{L}(\theta)
=
\lambda_u \sum_{k=0}^{N-1} w_k &\|\hat{u}_{2,k}-u_{2,k}\|_2^2
+
\lambda_x \sum_{k=0}^{N} w_k \|\hat{x}_{2,k}-x_{2,k}\|_2^2\\
&+
\lambda_g \sum_{k=0}^{N} \mathrm{ReLU}\!\left(d_{\mathrm{safe}}^2-\|p_{1,k}-\hat{p}_{2,k}\|_2^2\right)^2,
\end{aligned}
\end{equation}

where $\hat{p}_{2,k}$ is obtained from $\hat{x}_{2,k}$ via the Frenet-to-Cartesian mapping.
We use exponentially decaying weights $w_k=\gamma^k$ to emphasize near-term accuracy.

\paragraph{Dataset and hyperparameters.}
We train on $27{,}067$ trajectory samples with horizon $N=10$ and time step $dt=0.05$\,s.
The network structure and training configuration have been summarized in Appendix~B. Table~\ref{tab:br_mlp_metrics} reports validation metrics for the output of the network. 
We emphasize that the surrogate is an internal component used to instantiate $\widehat{\mathcal{B}}_2$ under asymmetric information, not a core contribution by itself.
\begin{table}[t]
\centering
\caption{Validation performance of $\widehat{\mathcal{B}}_2$.}
\label{tab:br_mlp_metrics}
\small
\setlength{\tabcolsep}{8pt}
\renewcommand{\arraystretch}{1.12}
\begin{tabular}{@{} l c @{}}
\toprule
\textbf{Metric} & \textbf{Value} \\
\midrule
Control RMSE (acceleration $a$)  [m/s$^2$]  & 0.155 \\
Control RMSE (steering $\delta$) [rad] & 0.073 \\
Rollout RMSE (heading angle $\psi$) [rad] & 0.0550 \\
Rollout RMSE (progress $s$) [m] & 0.0077 \\
Rollout RMSE (lateral error $t$) [m] & 0.0236 \\
Overall Cartesian position RMSE [m] & 0.0248 \\
Overall Contour RMSE [m] & 0.01561 \\
Max collision-constraint violation [m$^2$] & $3.31\times 10^{-2}$ \\
% Max collision-constraint violation\footnotemark\ [m$^2$] & $3.31\times 10^{-2}$ \\
\bottomrule
\end{tabular}
\end{table}

\subsection{Monte Carlo Setup}
\label{sec:exp_mc}

We evaluate over $1200$ randomly sampled initial conditions.
To focus on regimes where strategic interaction is nontrivial, we sample initial conditions such that the agents start within a proximity window ($\|p_{1,0}-p_{2,0}\|_2 \le 0.7$\,m), which activates collision coupling and matches the interaction distribution used to train $\widehat{\mathcal{B}}_2$.
All sampled initial conditions additionally satisfy track bounds and the initial collision constraint.

The best-response surrogate is trained on a disjoint set of trajectories (no overlap in initial conditions), and all reported test results are computed on held-out trials. We evaluate two planning pipelines:

\noindent\textbf{(i) Full-information equilibrium baselines.}
We run joint game solvers that assume online access to both players' objectives and constraints (including Player~2), and solve (or approximate) the coupled equilibrium conditions in \eqref{eq:joint_kkt_compact}. In our implementation, we use DGSQP and IBR as representative baselines.

\noindent\textbf{(ii) Proposed asymmetric reduced formulation (ours).}
We solve the reduced system in Sec.~\ref{sec:mcp_impl} using only Player~1's model $(J_1,h_1,g_1)$ and the embedded operator constraint $Z_2=\widehat{\mathcal{B}}_2(X_1)$. Player~2's objective is not queried online.

\noindent\textbf{Fairness.}
All methods use the same horizon $N$, time step $dt$, state/input bounds, collision constraint, and identical initialization and warm-start policy when applicable.

\noindent\textbf{Hardware \& Runtime.} 
All experiments were run on an Intel 13th Gen Core i9-13900HX system with 16\,GB RAM, running Windows 11.
We report wall-clock solve times for the online planning step only, excluding offline dataset generation and training of $\widehat{\mathcal{B}}_2$.
To reduce variability, we restrict numerical linear algebra and solver backends to a single CPU thread and report median and 95th-percentile runtime over Monte Carlo trials.

\begin{figure*}[t]
    \centering
    \includegraphics[width=\linewidth]{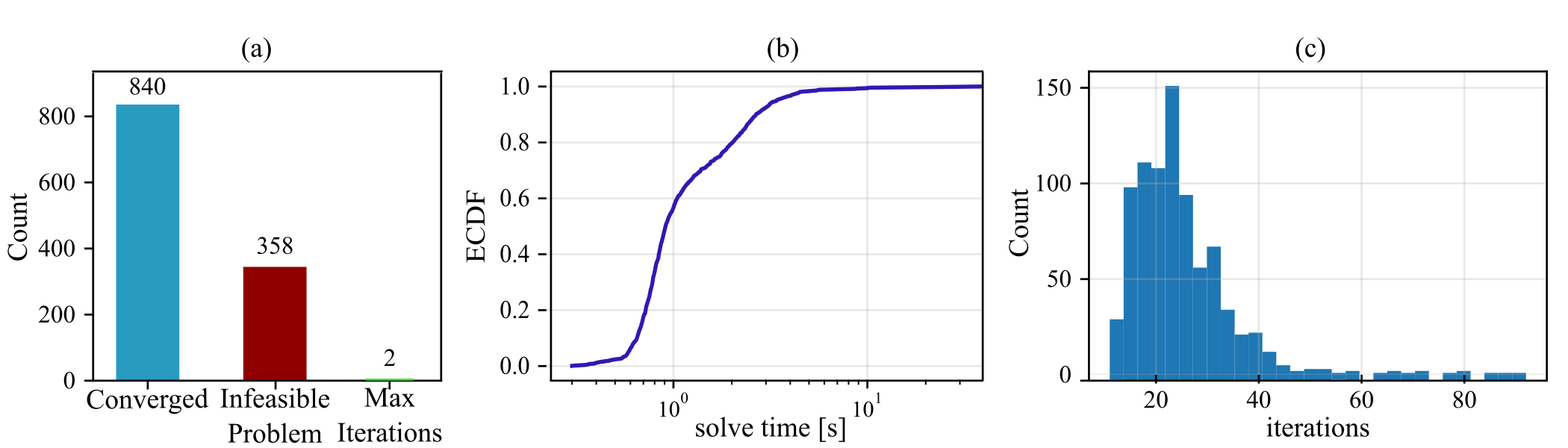}
    \caption{\textbf{Reduced formulation performance (ours) over $N=1200$ Monte Carlo instances.}
    (a) IPOPT termination status counts.
    (b) Empirical CDF (ECDF) of wall-clock solve time (log-scale x-axis).
    (c) Distribution of IPOPT iteration counts on converged runs (clipped at the 99th percentile for readability).}
    \label{fig:mcp_perf}
\end{figure*}

\begin{figure}[t]
    \centering
    \includegraphics[width=\linewidth]{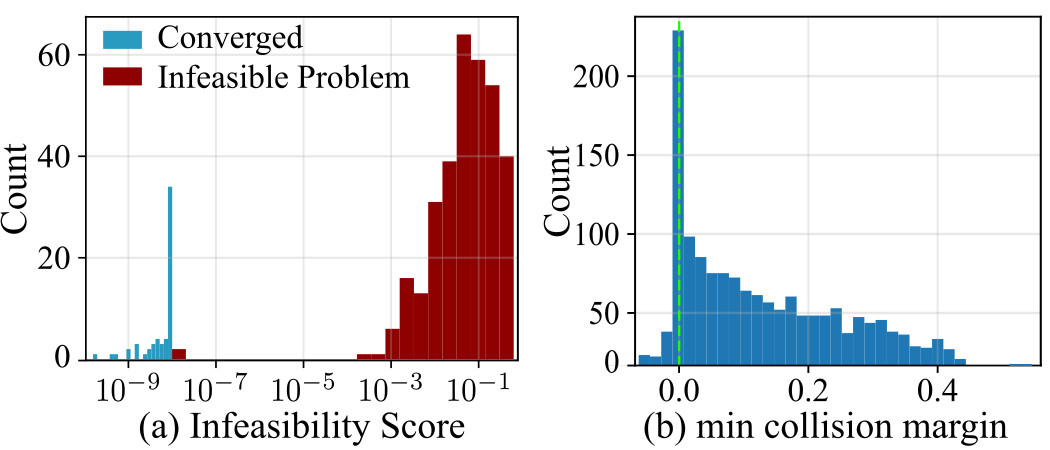}
    \caption{\textbf{Constraint diagnostics for the reduced formulation (ours).}
    (a) Aggregate infeasibility score $s_{\mathrm{infeas}}$ stratified by termination status (log-scale x-axis).
    (b) Minimum collision margin $\min_k(\|p_{1,k}-p_{2,k}\|_2-d_{\mathrm{safe}})$; values below zero indicate safety-distance violations.}
    \label{fig:mcp_constraints}
\end{figure}

% % \begin{figure}[t]
% % \centering
% % \subfloat[Infeasibility score]{
% %   \includegraphics[width=0.9\linewidth]{Figures/constraint_violations-a.png}
% %   \label{fig:mcp_infeas_score}
% % }\\
% % \subfloat[Min collision margin $(\mathrm{dist}-d_{\mathrm{safe}})$]{
% %   \includegraphics[width=0.9\linewidth]{Figures/constraint_violations-b.png}
% %   \label{fig:mcp_collision_margin}
% % }
% % \caption{Constraint satisfaction summary: (a) infeasibility score distribution, (b) minimum collision margin.}
% % \label{fig:constraint_violations}
% % \end{figure}

% These results validate the intended role of the proposed methodology.
% Even without online access to Player~2's objective and constraints, the reduced formulation produces feasible interaction plans on the majority of interacting test cases and does so with predictable iteration and runtime behavior.
% At the same time, the failure modes (infeasibility detection and occasional safety-margin violations) emphasize the central trade-off of the approach: the quality of the embedded best-response operator directly affects feasibility and safety, motivating careful dataset curation, out-of-distribution detection, and conservative safety handling when deploying $\widehat{\mathcal{B}}_2$.

\subsection{Results}
\label{sec:exp_results}

Fig.~\ref{fig:mcp_perf} summarizes the behavior of the proposed reduced formulation over $N=1200$ interaction-rich initial conditions.
Across this Monte Carlo set, IPOPT reports \texttt{Solve\_Succeeded} on $840/1200$ instances ($70.0\%$).
Among the non-success outcomes, the dominant termination mode is \texttt{Infeasible\_Problem\_Detected} ($358/1200$, $29.8\%$), while iteration-limit and restoration failures are rare ($2/1200$ total).
Together, these statistics indicate that most failures arise from (numerical) infeasibility of the reduced problem under the embedded response constraint, rather than slow convergence on otherwise feasible instances (Fig.~\ref{fig:mcp_perf}(a)).

Runtime behavior is summarized by the empirical CDF in Fig.~\ref{fig:mcp_perf}(b).
Conditioned on successful runs, solve time is tightly concentrated with a median below one second and a modest tail (see Table~\ref{tab:mc_summary} for median and p95).
The corresponding iteration distribution in Fig.~\ref{fig:mcp_perf}(c) shows that converged instances typically terminate within a few dozen IPOPT iterations, consistent with the observed runtime concentration.

To summarize feasibility across heterogeneous constraints, we report an aggregate infeasibility score
\begin{equation}
s_{\mathrm{infeas}} := \max\{e_{\mathrm{dyn}},\, e_{\mathrm{col}},\, e_{\mathrm{bnd}}\},
\end{equation}
where each $e_{\bullet}$ denotes excess violation beyond its corresponding tolerance (definitions and tolerances are provided in the Appendix).
Fig.~\ref{fig:mcp_constraints}(a) shows that successful runs cluster near numerical tolerances, whereas infeasibility-detected cases exhibit substantially larger violations.

Safety is evaluated by the minimum collision margin along the horizon,
$\min_{k}(\|p_{1,k}-p_{2,k}\|_2-d_{\mathrm{safe}})$.
Fig.~\ref{fig:mcp_constraints}(b) reports the distribution of this margin: most returned solutions maintain positive separation, but a nontrivial tail crosses below zero, indicating safety-distance violations for a subset of instances.
This behavior highlights the key trade-off in structural reduction: feasibility and safety depend directly on the accuracy and distributional validity of the embedded response operator.
In practice, this motivates conservative safety handling (e.g., tightened margins or safety buffers) and out-of-distribution detection when deploying $\widehat{\mathcal{B}}_2$.

\begin{figure*}[ht]
    \centering
    \includegraphics[width=\linewidth]{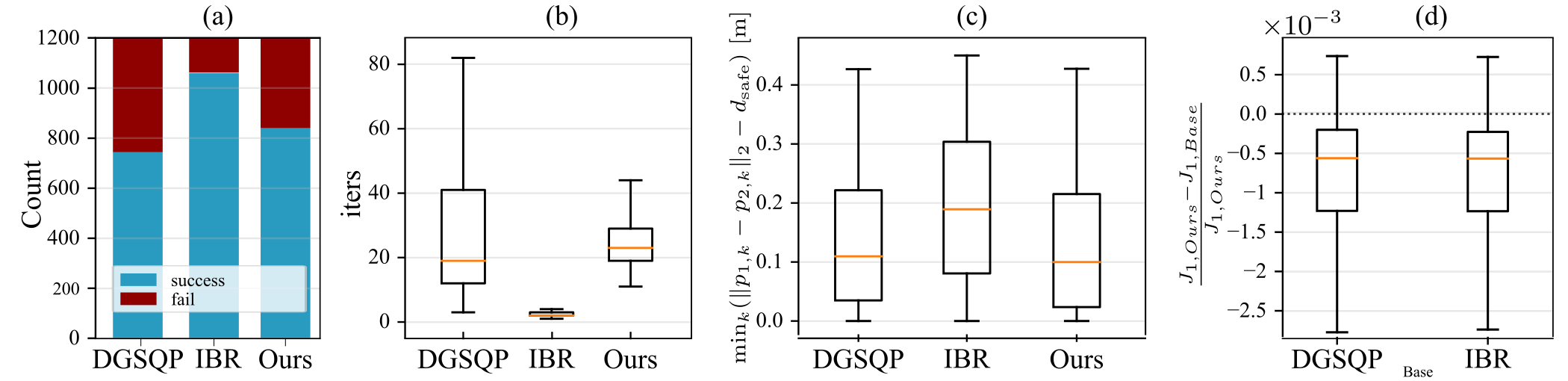}
    \caption{\textbf{Comparison to full-information baselines over $N=1200$ Monte Carlo instances.}
    (a) Iterations on successful runs.
    (b) Minimum collision margin on successful runs.
    (c) Paired ego-cost differences $\Delta J_1 := J_1^{\mathrm{ours}}-J_1^{\mathrm{base}}$ on instances where both solvers succeeded (negative is better for ours).
    (d) Outcome counts (success vs.\ non-success).}
    \label{fig:compare_panels}
\end{figure*}

\begin{table*}[ht]
\centering
% \caption{Monte Carlo summary on $1200$ interacting initial conditions. Reported times are success-only. Collision violation rate is the fraction of successful runs with $\min_k(\|p_{1,k}-p_{2,k}\|_2-d_{\mathrm{safe}}) < 0$.}
\caption{Monte Carlo summary on $1200$ interacting initial conditions. Reported times are success-only. Collision violation rate is an \emph{ex-post} safety check computed from the Cartesian separation margin $\min_k(\|p_{1,k}-p_{2,k}\|_2-d_{\mathrm{safe}}) < 0$.}

\label{tab:mc_summary}
\small
\setlength{\tabcolsep}{5.5pt}
\renewcommand{\arraystretch}{1.12}
\begin{tabular}{lcccccc}
\toprule
Solver & Success (\%) & Median time [s] & p95 time [s] & Median iters & p95 iters & Coll.\ viol.\ (\%) \\
\midrule
DGSQP & 62.0 & 0.697 & 7.337 & 19 & 255 & 0.81 \\
IBR   & 88.4 & 0.243 & 0.838 & 2  & 3   & 8.39 \\
Ours (Data--Driven $\widehat{\mathcal{B}}_2$) & 70.0 & 0.838 & 2.204 & 23 & 41 & 7.14 \\
\bottomrule
\end{tabular}
% \todo[inline]{Reconcile nonzero collision-violation rates with the success/feasibility definition. If collision margin is evaluated with a stricter post-check (different mapping/tolerance) than the solver constraints, state this explicitly and report both the solver constraint residual and the post-evaluation margin.}

\end{table*}
\subsection{Comparison to Full-Information Solvers}
\label{sec:exp_compare}

We compare the proposed asymmetric reduced formulation (ours) against full-information game solvers that assume online access to both players' objectives and constraints (including Player~2).
Table~\ref{tab:mc_summary} reports overall success rates, together with success-only runtime and iteration statistics on the same $N=1200$ Monte Carlo initial conditions.

Fig.~\ref{fig:compare_panels}(a) shows the success and failure rate distribution for each method.
IBR achieves the highest success rate and the lowest median runtime in this benchmark; however, it requires repeated online solution of Player~2's best-response subproblem and therefore relies on full access to Player~2's optimization model.
By contrast, our method eliminates any online dependence on Player~2's objective/constraints by enforcing the learned response feasibility constraint inside the NLP.
This introduces additional per-iteration work (network evaluation, rollout, and Jacobian contributions), and our current implementation does not yet exploit deployment-oriented accelerations such as code generation or compiled backends.
Despite these added components, our solver remains competitive and exhibits a substantially lighter runtime tail than DGSQP (Table~\ref{tab:mc_summary}), which is particularly important for predictable receding-horizon operation.

Fig.~\ref{fig:compare_panels}(b) compares iteration counts on successful runs.
IBR typically terminates in a few outer iterations, whereas DGSQP and our method often require tens of iterations, reflecting their different numerical structures and termination criteria.
Fig.~\ref{fig:compare_panels}(c) compares the minimum collision margin on successful runs.
In this benchmark, DGSQP yields the lowest ex-post collision-violation rate (Table~\ref{tab:mc_summary}), while IBR and our method exhibit higher violation rates under this margin check, motivating conservative safety buffers when deploying approximate response models.

Because the game is nonconvex and may admit multiple locally stationary equilibria, direct comparison of raw objective values across different instances is not meaningful.
Instead, we compare solution quality using paired, instance-matched differences on the subset where both methods succeed.
Let $\Delta J_1 := J_1^{\mathrm{ours}}-J_1^{\mathrm{base}}$, so $\Delta J_1<0$ indicates lower ego cost for our method on the same instance.
Fig.~\ref{fig:compare_panels}(d) shows that the paired $\Delta J_1$ distribution is centered near zero with a negative median against both baselines, while a small number of outliers yield positive $\Delta J_1$ due to convergence to different local equilibria, which is expected in nonconvex constrained games.

\section{Discussion}
\label{sec:discussion}

The experiments highlight the main contribution of this paper: Nash-consistent planning can be instantiated under asymmetric information by replacing the second player’s online optimality block with an offline-compiled response feasibility relation.
In the resulting reduced formulation, Player~1 remains explicitly optimal with respect to its own objective and constraints, while Player~2 enters only through the embedded operator constraint $Z_2=\widehat{\mathcal{B}}_2(X_1)$.
This structural reduction enables online planning without requiring access to Player~2’s objective or constraints at runtime, while remaining compatible with standard NLP machinery.

Full-information solvers act as oracles for equilibrium computation but require privileged access to both players’ optimization models.
In contrast, the proposed method achieves comparable paired ego costs on instances where both methods succeed, while operating under reduced information.
IBR is the fastest baseline in our benchmark, but it relies on repeated online best-response solves for Player~2, which are unavailable in the asymmetric-information setting.
DGSQP exhibits heavier runtime tails and a lower success rate in this setting, reflecting the sensitivity of coupled nonconvex equilibrium solves under tight collision constraints.

The dominant non-success outcome for the reduced formulation is infeasibility detection, and a subset of successful runs exhibit small safety-margin violations under ex-post evaluation.
These behaviors are consistent with the fact that general nonconvex dynamic games do not guarantee existence or uniqueness of a (generalized) Nash equilibrium, and with the reliance of the reduced system on an approximate embedded response model.
Practically, this suggests treating the response operator as a feasibility- and safety-critical component, motivating interaction-rich training data, out-of-distribution detection, and conservative safety handling (e.g., tightened margins or lightweight safety filters) when deploying $\widehat{\mathcal{B}}_2$.\\

\noindent\textbf{Limitations and future work.}
This work emphasizes structural reduction rather than solver engineering; our current implementation incurs additional per-iteration work from network evaluation and rollout, and does not yet exploit deployment-oriented accelerations such as code generation or compiled backends.
Moreover, the evaluation is open-loop on a fixed track and a kinematic bicycle model.
Future work will focus on (i) optimizing the implementation and exploiting structure for faster solves, (ii) improving the accuracy and optimization-friendliness of response representations, and (iii) extending to receding-horizon closed-loop execution with model mismatch and conservative safety mechanisms.

\section{Conclusion}
\label{sec:conclusion}

We presented an asymmetric structural reduction for two-player dynamic games that eliminates the need for online access to an opponent's objective and constraints.
The key idea is to replace the second player's optimality conditions with an offline-compiled response feasibility relation, instantiated either by a computed best-response operator or a learned surrogate $\widehat{\mathcal{B}}_2$.
This yields a reduced planning problem in which the ego agent remains explicitly optimal while the opponent is enforced through an embedded response constraint.

In a two-player racing benchmark with tight collision coupling, the proposed method produces feasible interaction plans on the majority of interacting test cases and achieves ego costs comparable to full-information equilibrium baselines on paired success instances, while requiring only ego model access online.
The observed failure modes emphasize the central trade-off of the approach: feasibility and safety depend on the quality and coverage of the embedded response operator.

\bibliographystyle{plainnat}
\bibliography{references}

@inproceedings{Soltanian20251419,
  title={PACE: A Framework for Learning and Control in Linear Incomplete-Information Differential Games},
  author={Soltanian, Seyed Yousef and Zhang, Wenlong},
  booktitle={7th Annual Learning for Dynamics$\backslash$\& Control Conference},
  pages={1419--1433},
  year={2025},
  organization={PMLR},
url={https://arxiv.org/abs/2504.17128}
}

@InProceedings{pmlrv283kim25a,
  title = 	 {Learning Two-agent Motion Planning Strategies from Generalized Nash Equilibrium for Model Predictive Control},
  author =       {Kim, Hansung and Zhu, Edward L. and Lim, Chang Seok and Borrelli, Francesco},
  booktitle = 	 {Proceedings of the 7th Annual Learning for Dynamics \&amp; Control Conference},
  pages = 	 {112--123},
  year = 	 {2025},
  volume = 	 {283},
  series = 	 {Proceedings of Machine Learning Research},
  month = 	 {04--06 Jun},
  publisher =    {PMLR},
  url = 	 {https://proceedings.mlr.press/v283/kim25a.html},
}

@InProceedings{pmlrv168espinoza22a,
  title = 	 {Deep Interactive Motion Prediction and Planning: Playing Games with Motion Prediction Models},
  author =       {Espinoza, Jose Luis Vazquez and Liniger, Alexander and Schwarting, Wilko and Rus, Daniela and Gool, Luc Van},
  booktitle = 	 {Proceedings of The 4th Annual Learning for Dynamics and Control Conference},
  pages = 	 {1006--1019},
  year = 	 {2022},
  volume = 	 {168},
  series = 	 {Proceedings of Machine Learning Research},
  month = 	 {23--24 Jun},
  publisher =    {PMLR},
  url = 	 {https://proceedings.mlr.press/v168/espinoza22a.html},
}

@inproceedings{Williams2023Distributed,
title = "Distributed Potential iLQR: Scalable Game-Theoretic Trajectory Planning for Multi-Agent Interactions",
author = "Zach Williams and Jushan Chen and Negar Mehr",
year = "2023",
doi = "10.1109/ICRA48891.2023.10161176",
language = "English (US)",
series = "Proceedings - IEEE International Conference on Robotics and Automation",
publisher = "Institute of Electrical and Electronics Engineers Inc.",
pages = "3476--3482",
booktitle = "Proceedings - ICRA 2023",
url={https://arxiv.org/abs/2303.04842v1}
}

@INPROCEEDINGS{RAPID2023Mehr,
  author={Jia, Yixuan and Bhatt, Maulik and Mehr, Negar},
  booktitle={2023 European Control Conference (ECC)}, 
  title={RAPID: Autonomous Multi-Agent Racing using Constrained Potential Dynamic Games}, 
  year={2023},
  volume={},
  number={},
  pages={1-8},
  doi={10.23919/ECC57647.2023.10178387},
url={https://doi.org/10.23919/ECC57647.2023.10178387}}

@INPROCEEDINGS{Wei2014CConstantSpeed,
  author={Wei, Junqing and Snider, Jarrod M. and Gu, Tianyu and Dolan, John M. and Litkouhi, Bakhtiar},
  booktitle={2014 IEEE Intelligent Vehicles Symposium Proceedings}, 
  title={A behavioral planning framework for autonomous driving}, 
  year={2014},
  volume={},
  number={},
  pages={458-464},
  doi={10.1109/IVS.2014.6856582},
url={https://doi.org/10.1109/IVS.2014.6856582}}

@article{chandra2025multi,
  title={Multi-robot navigation in social mini-games: Definitions, taxonomy, and algorithms},
  author={Chandra, Rohan and Singh, Shubham and Luo, Wenhao and Sycara, Katia},
  journal={arXiv preprint arXiv:2508.13459},
  year={2025},
url={https://arxiv.org/abs/2508.13459v3}
}

@article{smith2025mutual,
  title={Mutual Adaptation and Influence: Survey of Latent Dynamics Models in Human-Robot Interaction},
  author={Smith, Mason O and Amatya, Sunny and Soltanian, Seyed Yousef and Bush, Jonathan and Zhang, Wenlong},
  journal={Authorea Preprints},
  year={2025},
  publisher={Authorea},
url={https://doi.org/10.36227/techrxiv.175459948.88124600/v2}
}

@article{liu2023learning,
  title={Learning to play trajectory games against opponents with unknown objectives},
  author={Liu, Xinjie and Peters, Lasse and Alonso-Mora, Javier},
  journal={IEEE Robotics and Automation Letters},
  volume={8},
  number={7},
  pages={4139--4146},
  year={2023},
  publisher={IEEE},
url={https://arxiv.org/abs/2211.13779v3}
}

@article{Li2023SolvingSC,
  title={Solving Strongly Convex and Smooth Stackelberg Games Without Modeling the Follower},
  author={Yansong Li and Shuo Han},
  journal={2023 American Control Conference (ACC)},
  year={2023},
  pages={2332-2337},
  url={https://api.semanticscholar.org/CorpusID:257496190}
}

@article{sin2020iterativeBRDrawback,
  title={Iterative best response for multi-body asset-guarding games},
  author={Sin, Emmanuel and Arcak, Murat and Philbrick, Douglas and Seiler, Peter},
  journal={arXiv preprint arXiv:2011.01893},
  year={2020},
url={https://arxiv.org/abs/2011.01893v1}
}

@article{lidard2024KLGameBlending,
  title={Blending data-driven priors in dynamic games},
  author={Lidard, Justin and Hu, Haimin and Hancock, Asher and Zhang, Zixu and Contreras, Albert Gim{\'o} and Modi, Vikash and DeCastro, Jonathan and Gopinath, Deepak and Rosman, Guy and Leonard, Naomi Ehrich and others},
  journal={arXiv preprint arXiv:2402.14174},
  year={2024},
url={https://www.roboticsproceedings.org/rss20/p020.pdf},
}

@article{lopez2026data,
  title={On Data-based Nash Equilibria in LQ Nonzero-sum Differential Games},
  author={Lopez, Victor G and M{\"u}ller, Matthias A},
  journal={arXiv preprint arXiv:2601.11320},
  year={2026},
url={https://arxiv.org/abs/2601.11320v1}
}

@article{dockner2001coordinate,
  title={Coordinate transformations and derivation of open-loop Nash equilibria},
  author={Dockner, EJ and Leitmann, G},
  journal={Journal of Optimization Theory and Applications},
  volume={110},
  number={1},
  pages={1--15},
  year={2001},
  publisher={Springer},
url={https://doi.org/10.1023/A:1017511827387}
}

@inproceedings{SchwagerGameTheroreticPlanning,
author = {Wang, Mingyu and Wang, Zijian and Talbot, John and Gerdes, J. and Schwager, Mac},
year = {2019},
month = {06},
pages = {},
title = {Game Theoretic Planning for Self-Driving Cars in Competitive Scenarios},
  booktitle = {Proceedings of Robotics: Science and Systems (RSS)},
doi = {10.15607/RSS.2019.XV.048},
  url={https://api.semanticscholar.org/CorpusID:197466062}
}

@inproceedings{fridovich2020efficient,
  title={Efficient iterative linear-quadratic approximations for nonlinear multi-player general-sum differential games},
  author={Fridovich-Keil, David and Ratner, Ellis and Peters, Lasse and Dragan, Anca D and Tomlin, Claire J},
  booktitle={2020 IEEE international conference on robotics and automation (ICRA)},
  pages={1475--1481},
  year={2020},
  organization={IEEE},
  DOI={10.1109/ICRA40945.2020.9197129},
  url={https://doi.org/10.1109/ICRA40945.2020.9197129}
}

@article{le2022algames,
  title={Algames: a fast augmented lagrangian solver for constrained dynamic games},
  author={Le Cleac’h, Simon and Schwager, Mac and Manchester, Zachary},
  journal={Autonomous Robots},
  volume={46},
  number={1},
  pages={201--215},
  year={2022},
  publisher={Springer},
  URL={https://doi.org/10.1007/s10514-021-10024-7},
  DOI={10.1007/s10514-021-10024-7},
}

@INPROCEEDINGS{zhu2024sequentialdgsqp,
  author={Zhu, Edward L. and Borrelli, Francesco},
  booktitle={2023 IEEE International Conference on Robotics and Automation (ICRA)}, 
  title={A Sequential Quadratic Programming Approach to the Solution of Open-Loop Generalized Nash Equilibria}, 
  year={2023},
  volume={},
  number={},
  pages={3211-3217},
  doi={10.1109/ICRA48891.2023.10160799},
  URL={https://doi.org/10.1109/ICRA48891.2023.10160799},
}

@article{standardGNEPSolution,
author = {Dreves, Axel and Facchinei, Francisco and Kanzow, Christian and Sagratella, Simone},
title = {On the solution of the KKT conditions of generalized Nash equilibrium problems},
journal = {SIAM Journal on Optimization},
volume = {21},
number = {3},
pages = {1082-1108},
year = {2011},
doi = {10.1137/100817000},
URL = { https://doi.org/10.1137/100817000},
eprint = {https://doi.org/10.1137/100817000},
}

@inbook{Noncooperative1999Tamer,
title = {Noncooperative Finite Games: N-Person Nonzero-Sum},
booktitle = {Dynamic Noncooperative Game Theory, 2nd Edition},
chapter = {3},
pages = {77-160},
doi = {10.1137/1.9781611971132.ch3},
URL = {https://epubs.siam.org/doi/abs/10.1137/1.9781611971132.ch3},
eprint = {https://epubs.siam.org/doi/pdf/10.1137/1.9781611971132.ch3},
author={Ba{\c{s}}ar, Tamer and Olsder, Geert Jan},
year={1998},
publisher={SIAM},
}

@article{pustilnik2025generalized,
  title={Generalized Nash Equilibrium Solutions in Dynamic Games With Shared Constraints},
  author={Pustilnik, Mark and Borrelli, Francesco},
  journal={arXiv preprint arXiv:2502.19569},
  year={2025},
URL = {https://arxiv.org/abs/2502.19569},
}

\appendix
% ============================================================
\section*{Appendix A: Track Parametrization \& Bounds}
\label{app:model_params}

We consider a constant-curvature planar track parameterized by arc-length $s$ and lateral offset $t$ relative to the centerline. The centerline is a circular arc of radius $R$ with curvature $\kappa=1/R$. The continuous-time kinematic bicycle model in the Frenet frame used in the implementation is
\begin{equation}
    \begin{aligned}
        \dot v &= a, \\
\dot s &= \frac{v\cos(\psi+\beta)}{1-\kappa t}, \\
\dot t &= v\sin(\psi+\beta), \\
\dot \psi &= \frac{v\sin\beta}{l_r} \;-\; \frac{\kappa v\cos(\psi+\beta)}{1-\kappa t},
    \end{aligned}
\end{equation}
where $\beta(\delta) = \arctan\!\Big(\frac{l_f}{l_f+l_r}\tan\delta\Big)$.
We discretize using forward Euler with step size $dt$ to obtain $x_{k+1}=f(x_k,u_k)$. 

\subsection*{A.1 Frenet-to-Cartesian Mapping (quarter-circle track)}
For a constant-curvature track with radius $R$, define the centerline heading angle
\begin{equation}
\theta(s) := \frac{s}{R}.
\end{equation}
The Cartesian centerline corresponding to $(s,0)$ is
\begin{equation}
p_c(s)=\begin{bmatrix}X_c(s)\\Y_c(s)\end{bmatrix}
=\begin{bmatrix}R\sin\theta(s)\\R\big(1-\cos\theta(s)\big)\end{bmatrix}.
\end{equation}
A lateral offset $t$ is applied along the track normal, yielding the Cartesian position for a Frenet point $(s,t)$:
\begin{equation}
p(s,t)=\begin{bmatrix}X(s,t)\\Y(s,t)\end{bmatrix}
=\begin{bmatrix}
R\sin\theta(s)-t\sin\theta(s) \\
R\big(1-\cos\theta(s)\big)+t\cos\theta(s)
\end{bmatrix}.
\label{eq:frenet_to_xy}
\end{equation}
Collision checking in the experiments is performed in Cartesian space using \eqref{eq:frenet_to_xy}.

Table~\ref{tab:app_track_params} reports the track and vehicle parameters used in all experiments. Table~\ref{tab:app_bounds_params} lists the state/input bounds and safety distance. The progress coordinate is limited to a quarter-circle arc-length,
\begin{equation}
s \in [0,\, R\pi/2] = [0,\, 5.4978]\ \mathrm{m}.
\end{equation}

\begin{table}[t]
\centering
\caption{Track and kinematic bicycle parameters (shared by both agents).}
\label{tab:app_track_params}
\small
\setlength{\tabcolsep}{7pt}
\renewcommand{\arraystretch}{1.12}
\begin{tabular}{l l c}
\toprule
\textbf{Symbol} & \textbf{Description} & \textbf{Value} \\
\midrule
$R$ & Track radius & $3.5\ \mathrm{m}$ \\
$\kappa$ & Track curvature & $1/R = 0.2857\ \mathrm{m}^{-1}$ \\
$dt$ & Discrete-time step (Euler) & $0.05\ \mathrm{s}$ \\
$l_f$ & Front axle distance & $0.13\ \mathrm{m}$ \\
$l_r$ & Rear axle distance & $0.13\ \mathrm{m}$ \\
$L$ & Wheelbase & $l_f+l_r = 0.26\ \mathrm{m}$ \\
$R\pi/2$ & Quarter-circle arc-length & $5.4978\ \mathrm{m}$ \\
\bottomrule
\end{tabular}
\end{table}

\begin{table}[t]
\centering
\caption{State and input bounds and collision safety distance used in the experiments.}
\label{tab:app_bounds_params}
\small
\setlength{\tabcolsep}{6pt}
\renewcommand{\arraystretch}{1.12}
\begin{tabular}{l c c}
\toprule
\textbf{Parameter} & \textbf{Lower bound} & \textbf{Upper bound} \\
\midrule
$v$ [m/s] & $0$ & $2$ \\
$\psi$ [deg] & $-180$ & $180$ \\
$s$ [m] & $0$ & $R\pi/2$ \\
$t$ [m] & $-0.5$ & $0.5$ \\
\midrule
$a$ [m/s$^2$] & $-2$ & $2$ \\
$\delta$ [deg] & $-25$ & $25$ \\
\midrule
$d_{\mathrm{safe}}$ [m] & \multicolumn{2}{c}{$0.25$} \\
\bottomrule
\end{tabular}
\end{table}

% ============================================================
\section*{Appendix B: Best-Response Surrogate $\widehat{\mathcal{B}}_2$}
\label{app:br_surrogate}

This appendix documents the learned best-response surrogate $\widehat{\mathcal{B}}_2$ used to instantiate the response-feasibility constraint in the reduced formulation. The surrogate is trained offline and queried online to produce a horizon-length control sequence for Player~2, which is then rolled out through the known kinematic bicycle model to obtain a dynamically feasible opponent trajectory.

The surrogate is implemented as a multi-layer perceptron (MLP) with $\tanh$ nonlinearities and is trained using AdamW. Table~\ref{tab:br_mlp_summary} reports the architecture and training hyperparameters used in all experiments.

% ---------- Table: BR surrogate key hyperparameters ----------
\begin{table}[b]
\centering
\caption{Best-response surrogate $\widehat{\mathcal{B}}_2$ (MLP) and training hyperparameters.}
\label{tab:br_mlp_summary}
\small
\setlength{\tabcolsep}{8pt}
\renewcommand{\arraystretch}{1.12}
\begin{tabular}{@{} l c @{}}
\toprule
\textbf{Hyperparameter} & \textbf{Value} \\
\midrule
Horizon $N$ & 10 \\
Time step $dt$ & 0.05 s \\
Hidden widths & [128, 128, 64] \\
Nonlinearity & $\tanh$ \\
Optimizer & AdamW \\
Learning rate & $3\times 10^{-4}$ \\
Weight decay & $1\times 10^{-4}$ \\
Batch size & 256 \\
Epochs & 500 \\
Grad clip (max-norm) & 5.0 \\
Time weights $w_k=\gamma^k$ & $\gamma=0.92$ \\
Loss weights $(\lambda_u,\lambda_x,\lambda_g)$ & (30, 150, 50) \\
Safety distance $d_{\mathrm{safe}}$ & 0.25 m \\
Train/val split & 80/20 \\
Total \# of samples $M$ & 27{,}067 \\
\bottomrule
\end{tabular}
\end{table}

% ============================================================
\section*{Appendix C: Tolerances \& Infeasibility Score Definition}
\label{app:tolerances_infeas}

All Monte Carlo experiments use a unified numerical tolerance \(\varepsilon=10^{-6}\) across solvers. 
To summarize feasibility and safety across heterogeneous constraints, we report solver-agnostic ex-post residuals computed from returned trajectories.
Let \(X_{1,k}\) and \(U_{1,k}\) denote Player~1's state and input, and let \(p_{i,k}\in\mathbb{R}^2\) denote the Cartesian position of agent \(i\) obtained by the Frenet-to-Cartesian mapping described in Appendix~\ref{app:model_params}.
We define:

\paragraph{Dynamics defect.}
The maximum one-step dynamics residual (infinity norm) is
\begin{equation}
e_{\mathrm{dyn}}
:= \max_{k=0,\dots,N-1} \left\|X_{1,k+1}-f(X_{1,k},U_{1,k})\right\|_{\infty},
\label{eq:app_dyn_defect}
\end{equation}
where \(f(\cdot)\) is the discrete-time augmented Frenet dynamics used in the solver.

\paragraph{Collision-avoidance violation.}
Collision avoidance is enforced in squared-distance form, \(\|p_{1,k}-p_{2,k}\|_2^2 \ge d_{\mathrm{safe}}^2\). 
We report both the minimum margin and the corresponding maximum violation:
\begin{align}
m_{\mathrm{col}} 
&:= \min_{k=0,\dots,N} \left(\|p_{1,k}-p_{2,k}\|_2^2 - d_{\mathrm{safe}}^2\right), \\
e_{\mathrm{col}}
&:= \max_{k=0,\dots,N}\max\!\left(0,\; d_{\mathrm{safe}}^2-\|p_{1,k}-p_{2,k}\|_2^2\right).
\label{eq:app_collision_violation}
\end{align}
Thus, \(m_{\mathrm{col}}\ge 0\) indicates satisfaction of the safety constraint, while \(e_{\mathrm{col}}>0\) quantifies the worst squared-distance violation.

\paragraph{Box-bound violation.}
For any variable \(z\) with componentwise bounds \(z_{\min}\le z \le z_{\max}\), we define the maximum bound violation as
\begin{equation}
e_{\mathrm{bnd}}(z)
:= \max\!\left\{0,\; \|z_{\min}-z\|_{\infty},\; \|z-z_{\max}\|_{\infty}\right\}.
\label{eq:app_bound_violation}
\end{equation}
We evaluate this quantity for the ego augmented state and controls, and (when applicable) the opponent state and controls, and report the maximum across these categories as \(e_{\mathrm{bnd}}\).

Finally, we combine these heterogeneous diagnostics into a single infeasibility score
\begin{equation}
s_{\mathrm{infeas}} := \max\{e_{\mathrm{dyn}},\, e_{\mathrm{col}},\, e_{\mathrm{bnd}}\}.
\label{eq:app_infeas_score}
\end{equation}
By construction, \(s_{\mathrm{infeas}}=0\) indicates that all tracked constraints are satisfied within tolerance, while larger values indicate the dominant excess violation among dynamics, collision avoidance, or bounds.

% ============================================================
\section*{Appendix D: Additional Diagnostics \& Qualitative Examples}
\label{app:additional_figs}

This appendix provides supplementary visual diagnostics that complement the quantitative Monte Carlo results in the main paper. 
We include (i) a visualization of solver success/failure across the sampled initial-condition distribution to illustrate where failures tend to occur in the state space, and (ii) representative open-loop trajectory pairs produced by the proposed method to qualitatively verify typical interaction patterns and constraint satisfaction on the track.
These figures are intended as supporting evidence and do not introduce additional claims beyond the main experimental section.

\begin{figure}[ht]
    \centering
    \includegraphics[width=\linewidth]{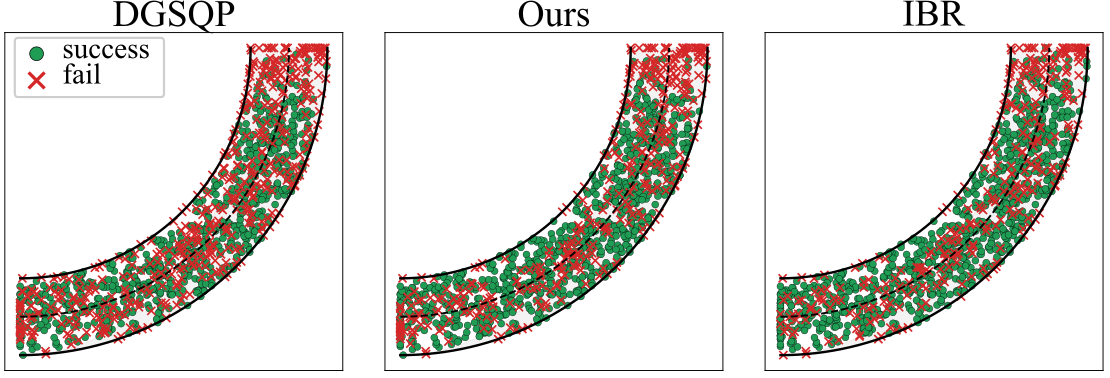}
    \caption{Solver outcomes across the distribution of initial states used in the benchmark. Each marker corresponds to one Monte Carlo initial-condition pair.}
    \label{fig:mc_convergence_map}
\end{figure}

\begin{figure}[ht]
    \centering
    \includegraphics[width=\linewidth]{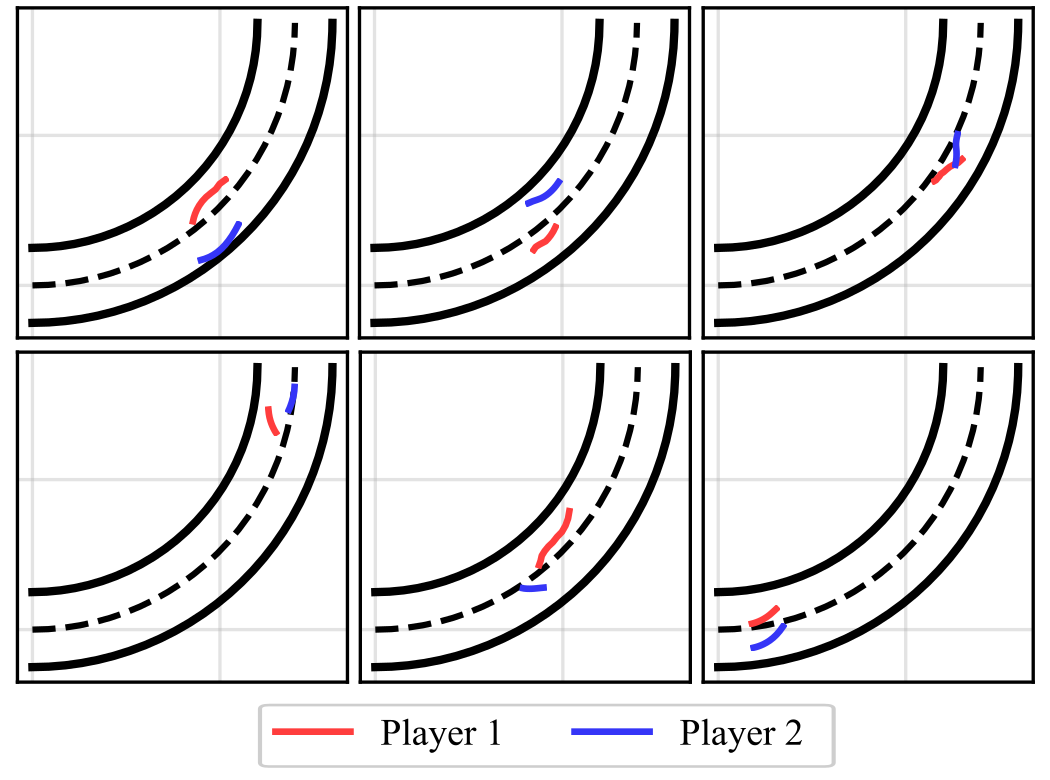}
    \caption{Representative open-loop trajectory pairs generated by the proposed method for multiple initial-condition pairs. Player~1 is shown in red and Player~2 in blue. }
    \label{fig:mcp_traj_samples}
\end{figure}

\end{document}